\documentclass[journal abbreviation, manuscript]{copernicus}

\usepackage{amsthm}
\usepackage[capitalise]{cleveref}
\usepackage{booktabs}
\usepackage{subcaption}
\usepackage{derivative}

\begin{document}

\title{Investigating the Relationship between Simulation Parameters and Flow Variables in Simulating Atmospheric Gravity Waves in Wind Energy Applications}


\Author[1]{Mehtab Ahmed}{Khan}
\Author[1]{Dries}{Allaerts}
\Author[1]{Simon}{Watson}
\Author[2]{Matthew}{Churchfield}

\affil[1]{Technical University Delft, Netherlands}
\affil[2]{National Renewable Energy Laboratory, USA}




\correspondence{Mehtab Ahmad Khan (m.a.khan-2@tudelft.nl)}

\runningtitle{TEXT}

\runningauthor{TEXT}

\received{}
\pubdiscuss{} 
\revised{}
\accepted{}
\published{}


\firstpage{1}

\maketitle

\begin{abstract}
Wind farms, particularly offshore clusters, are becoming larger than ever before. Besides influencing wind farms and local meteorology downstream, large wind farms can trigger atmospheric gravity waves in the inversion layer and the free atmosphere aloft. Wind farm-induced gravity waves can cause adverse pressure gradients upstream of the wind farm, that contribute to the global blockage effect, and favorable pressure gradients above and downstream of the wind farm that enhance wake recovery. 

Numerical modeling is a powerful means of studying wind farm-induced atmospheric gravity waves, but it comes with the challenge of handling spurious reflections of these waves from domain boundaries. Approaches like radiation boundary conditions and forcing zones are used to avoid the reflections. However, the simulation setup heavily relies on ad-hoc processes. For instance, the widely used Rayleigh damping method requires ad-hoc tuning to acquire a setup only applicable to a particular case. To surmount this hurdle, we conduct a systematic LES study for flow over a 2D hill and through wind farm canopies that explores the dependence of domain size and damping layer setup on parameters driving linearly stratified atmospheric flows. 

 Mainly the internal waves in the free atmosphere reflect from the boundaries, therefore by simulation linearly stratified conditions we focus on internal  waves only. The Froude number drives most of the internal wave properties, such as wavelengths, amplitude, and direction. Therefore, the domain sizing and Rayleigh damping layer setup mainly depends on the Froude number. We anticipated the effective wavelengths to be the correct length scale to size the domain and damping layer thickness. Also, the damping coefficient is scaled with Brunt-V\"ais\"al\"a frequency. 
 
 Using Froude numbers seen in wind farm applications, we propose some recommendations to limit the reflections to less than $10 \%$. Typically damping is done at the top boundary, but given the non-periodic lateral boundary conditions of practical wind farm simulation domains, we find that damping the inflow-outflow boundaries is of equal importance to the top boundary. The damping coefficient normalized with the Brunt-V\"ais\"al\"a frequency, buoyancy time scale, should be $1$ to $10$. The damping thickness should be at least one effective vertical wavelength; damping layers exceeding $1.5$ times the vertical wavelength could be redundant. The domain length should accommodate at least one effective horizontal wave. Likewise, the non-damped domain height should be at least one effective vertical wavelength. Moreover, Rayleigh damping does not dampen the waves completely, and the non-damped energy might accumulate over the simulation time.  
\end{abstract}

\introduction
\label{sec:intro}
 
 The size of a modern wind farm, especially offshore, can extend to several tens of kilometers, involving flow interactions on a regional scale and to great elevations into the free atmosphere. The energy and momentum extraction caused by a large wind farm is significant enough to decelerate the flow in the atmospheric boundary layer (ABL) \citep{Frandsen1992}, \citep{Smith2010}, which slowly recovers because of turbulent momentum transfer and the interplay of the large-scale driving pressure gradient and Coriolis effects. This is evident from the compound wake of large wind farms, which compared to small wind farm wakes, extend far beyond the wind farm in the streamwise direction. Generally, researchers study wind farm interactions with the ABL alone and not with the free atmosphere above, but a full understanding of large wind farm behavior requires an understanding of the farm's effect on the free atmosphere and vice versa.

The stability and temperature structure of both the ABL and the free atmosphere is deeply involved in the flow dynamics of large wind farms. For instance, the thickness of the ABL can significantly decrease during stable atmospheric conditions. In offshore environments in particular, the capping inversion, the relatively thin strongly stable layer that often forms between the ABL and the free atmosphere, can drop to a few hundred meters at night. In such conditions, wind farms can induce atmospheric gravity waves (AGWs), which include trapped gravity waves (TGWs) in the capping inversion and internal gravity waves (IGWs) in the free atmosphere aloft. Wind farm-induced AGWs were identified by \citet{Smith2010}, who analogized wind farms as semi-permeable obstacles that can deflect the flow upwards to displace the capping inversion, resulting in these buoyancy-driven waves. 

Wind farm-induced AGWs can affect wind farm performance by creating streamwise pressure gradients that accelerate or decelerate the flow into and through the farm, contributing to the phenomenon of "large wind-farm blockage" and affecting the way the compound wind-farm wake recovers. \cite{Smith2010}, \cite{Allaerts2018}, and \cite{Lanzilao2021} believe that AGWs cause the large wind-farm blockage effect at the entrance of a wind farm.  \citet{Wu2017} theorize that the large wind-farm blockage effect is the combined effect of AGWs and the cumulative turbine induction. The term ‘regional efficiency’ of wind farms was introduced by \citet{Allaerts2018a} to characterize the impact of atmospheric gravity waves on wind-farm performance. They claim that regional efficiency is of the same order wake-induced wind-farm efficiency losses. \citet{Allaerts2018} estimated a  $4$ to $6\%$ decrease in the annual energy production of an offshore wind farm caused by the global blockage created by AGWs. On the other hand, \citep{Lanzilao2022} and \citep{StipaEtAl2023a} argue that AGWs can also assist wind farm wake recovery because they can create a local favorable pressure gradient at the far end of and behind the wind farm. However, the extent of the favorable pressure and its positive effects on farm wake recovery are still unknown. Because of the sheer scale of AGWs and their interaction with wind farms, little if any experimental data exists to quantify their effects on wind farms; most investigations about wind-farm-induced AGWs are numerical.

Numerical flow simulation, particularly large-eddy simulation (LES), is currently a powerful approach to study AGWs. Gravity waves induced by large wind farms were first investigated with LES by \citet{Allaerts2017}. Since then, the major work on wind farm-induced gravity waves using LES was performed by \citet{2016_Allaerts_phdthesis}, \citet{Allaerts2015, Allaerts2017, Allaerts2018a}, \citet{Allaerts2018}, and \citet{Lanzilao2021, Lanzilao2022} using pseudospectral codes.  \citet{StipaEtAl2023a} and \citet{wes-2022-63} have also studied wind-farm-induced gravity wave effects using finite-volume codes.  

The simulation of wind-farm-induced atmospheric gravity waves on a finite domain requires special treatment at all domain boundaries other than the ground to stop the gravity waves from spuriously reflecting off of these boundaries that do not exist in reality.  AGWs propagate in all directions, so they will spuriously interact with the inflow, outflow, and top boundaries (all of which do not exist in reality).  Two main classes of methods exist to mitigate spurious AGW-domain boundary interaction: 1) "radiative" boundary conditions and 2) damping layers.

Radiative boundary conditions often take the form
\begin{equation}
    \frac{\partial \phi}{\partial t} + c_j \frac{\partial \phi}{\partial n_j} = 0
\end{equation}

where $\phi$ is the quantity being transported through a boundary with normal, $n_j$, and $c_j$ is wave transport velocity.  One way to imagine this boundary condition is that it attempts to set the transported quantity $\phi$ exactly to the value the wave causes it to be as the wave propagates to the boundary.  The difficulty is determining $c_j$ or a good approximate of it.  This is often only possible in idealized situations.  The radiative boundary condition is capable of allowing wave energy out of the domain only if the waves are moving perfectly normal to the boundary \citet{Durran1999}. However, the waves are typically moving at an angle relative to horizontal.  

A popular alternative to the radiative boundary condition is the damping layer.  Damping layers, sometimes referred to as "sponge zones," attempt to attenuate waves as the wave moves through the zone.  These zones are placed adjacent to domain boundaries and have a prescribed thickness.  A downside to such zones is that because they have thickness, one has to increase the domain size to account for their thickness, which adds some extra cost.  Typically damping layers are either of the viscous type or of the relaxation type, also known as Rayleigh damping layers (RDLs).  RDLs appear much more commonly used than viscous damping, and in our experience, RDLs are more effective than viscous damping.  Also, a recent study by \citet{Lanzilao2022} found that RDLs outperform radiative boundary conditions in minimizing AGW reflections in wind farm simulations.  Therefore, we will focus on the RDL in this study.

In principle, the quantity of interest, usually velocity but also sometimes temperature, entering an RDL is relaxed towards a prescribed reference value at a specified time scale as the wave travels through the RDL, reflects off the boundary, and travels through the RDL once again. Rayleigh damping enters the transport equations as a forcing term which takes the form
\begin{equation}
    f^{RDL} = -\frac{1}{\tau}c[\phi - \phi_{ref}]
\end{equation}
where $\phi$ is the field to be damped within the RDL.  For example, commonly the velocity field is damped, so $\phi$ = $U_i$ is driven towards $U_{{i}_{ref}}$, which could be defined as components of geostrophic wind.  Sometimes the situation is ageostrophic, and a reference horizontal velocity is not convenient, so only the vertical component of velocity is damped. The Rayleigh function, $c$, is the a critical part of RBL that ensures that the wave gradually dissipates energy as it travels through the RDL thickness. $c$ can be a linear, exponential, polynomial, or cosine function.  We use the cosine form given here for the upper boundary,
\begin{equation}
    c = [1- \cos(\frac{\pi}{s_\mathrm{ra}}{\frac{z - (L_\mathrm{z} - L_\mathrm{d})}{L_\mathrm{d}}})]
\end{equation}
where $z$ is the total height of the domain, $L_\mathrm{z}$ is the height of the non-damped region, $L_\mathrm{d}$ is the thickness of the damping layer, and $s_\mathrm{ra}$ is the damping fraction. The time scale, $\tau$, or damping coefficient controls how fast the quantity is relaxed toward a reference value. In a recent study by \citet{Lanzilao2022}, $1/\tau$ is scaled with the Brunt-V\"ais\"al\"a frequency (N) in pursuit of the optimal value for wind farm simulations, suggesting the value of $\xi = 1/\tau N$ to be 3. Whereas \citet{2016_Allaerts_phdthesis} tuned their RDL for optimal damping coefficient and used $1/\tau = 0.0001~s^-1$, translating into $\xi = 0.017$; they used the same value in most of their later studies including, \citet{Allaerts2017, Allaerts2018, Allaerts2018a}. The damping layer thickness is another essential RDL parameter as it determines the space available for the forcing to dissipate the incoming wave energy. \citet{klemp-Lilly1978} suggested the thickness to be more than one vertical wavelength of the wave. More recent studies also follow this convention but without any further investigation; \citet{2016_Allaerts_phdthesis}, \citet{Allaerts2017}, \citet{Allaerts2018}, and \citet{Lanzilao2021, Lanzilao2022} use a $15~km$ RDL only at the top boundary; the vertical wavelengths in their simulations were always less than $15~km$. 

The damping fraction $s_\mathrm{ra}$ was considered one of the two most critical parameters by \citet{Lanzilao2022}, which determines the fraction of the damping layer where a cosine function is applied. If $s_\mathrm{ra}$ is $0.8$, the cosine ramp up/down of damping is used for $80\%$ of the layer, and damping strength in the remaining $20\%$ of the RDL is maximum. The damping function type is another aspect, which \citet{Peric} found less significant while investigating the approaches to dampen internal waves. We also think the damping fraction and type of damping function have little impact on effective damping.  Instead, the damping coefficient and thickness (collectively referred to as damping characteristics in this study) could be the most important RDL parameters. More importantly, no consensus and recommendations on properly tuning RDL are found in the literature. 

As mentioned above, many of the existing LES studies of wind farm interaction with AGWs are performed with codes that are horizontally pseudospectral.  Psuedospectral codes have the advantage of nearly exponential error convergence with increasing spatial resolution, but they come with the limitation that all lateral domain boundaries must be periodic.  This poses a challenge for wind farm simulations because we desire to feed turbulent ABL flow into the domain and allow that flow with the wind turbine wakes to exit the domain.  Periodic boundaries normally would cause the wake flow exiting the domain to reenter on the inflow side.  The solution to this problem is the introduction of a forcing fringe region adjacent to the downstream boundary which forces the waked solution back toward the precomputed or concurrently computed pure turbulent ABL inflow solution \citep{InoueEtAl2014}.  Finite-volume, finite-difference, and finite-element methods do not have this requirement of lateral periodicity and hence do not need a forcing fringe.  Inflow data is simply fed into the domain on the inflow boundary with a Dirichlet condition and the outflow boundary is often some sort of Neumann or advection boundary condition that lets the waked flow out.  There is no need for a fringe zone that forces the wake solution back to the inflow solution.  

We mention all of this because it is important to note that pseudospectral codes with their periodic lateral boundary conditions effectively have no lateral boundaries.  RDLs only have to be applied adjacent to the top boundary.  Furthermore, the forcing fringe used to force the downstream flow back toward the desired inflow state is a form of Rayleigh damping.  This means that AGWs will not reflect off of the lateral boundaries in pseudospectral LES, they will simply exit the domain on one side and reenter on the opposite side.  However, the forcing fringe will have some effect on damping their lateral progression or may trigger spurious gravity waves, requiring an additional treatment\citep{Lanzilao2021}.  

\begin{figure}[h]
    \centering
    \includegraphics[width=0.9\textwidth]{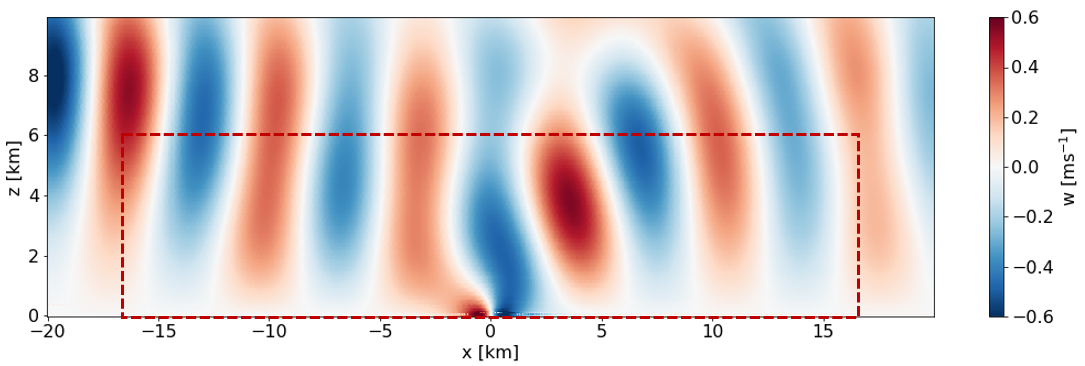}
    \caption{Vertical velocity contour of flow over a small hill located at x = 0 km. The region outside the red dotted box is Rayleigh damping layers at the inlet (left), top, and outlet (right).}
    \label{fig:ref_vc}
\end{figure}

The more common finite-volume/difference/element codes have real lateral boundary conditions that gravity waves will either reflect off of or spuriously interact with.  Therefore, RDLs must be placed adjacent to these boundaries as well as adjacent to the top boundary.  Very little guidance is given in the literature for gravity wave treatment for lateral boundaries in finite-volume/difference/element codes. \citet{StipaEtAl2023a} and \citet{wes-2022-63} used FVCs for their large wind farm LES which are not restricted to horizontally periodic boundary conditions. The latter used true inflow and outflow boundary conditions, which is complicated when modelling AGWs. Any inconsistency between the inflow and the internal fields, triggers non-physical waves at the inlet which propagate into the domain over time. Besides spurious waves, reflection of gravity waves from the domain boundaries is a huge challenge. As shown in \Cref{fig:ref_vc}, the gravity waves triggered by a small hill under linearly stratified conditions should move up and out of the numerical domain. But the reality of a finite domain with boundary conditions that do not exactly depict the actual physical conditions at the boundaries cause these waves to reflect. The non-dissipated wave energy accumulates at the boundaries and propagates back into the domain because it never leaves the domain, eventually making the solution non-physical. 

These current approaches to setting up RDL in wind farm simulations are all ad-hoc. Extensive fine-tuning of the damping parameters is required to set up a working simulation applicable only to a specific case \citep{Lanzilao2022}. Thus, setting up reflection-free simulations becomes tedious, computationally expensive, and time-consuming. One way to make the process less ad-hoc is to investigate possible relations of simulation setup with physical parameters driving flows through wind farms under stable atmospheric conditions.

This study aims to propose a baseline to set up simulations involving internal atmospheric gravity waves. The focus is identifying flow parameters dictating internal wave properties such as wavelength and direction. We anticipate that appropriate domain length, height, and damping layer thickness are related to the effective wavelengths, and the damping coefficient relates to the buoyancy frequency. By investigating the link between the internal wave properties and the simulation setup, we aim to propose recommendations for wind farm simulation set up under stable atmospheric conditions. As a starting point and for simplicity, we consider only the free atmosphere with a range of Froude Number (Fr) practical to wind farm applications. Investigating the free atmosphere separately is vital to handling the reflections, as the AGWs reflect off the boundaries. To this end, we consider two flow scenarios: first, we simulate flow over a bell-shaped 2D hill; second through wind farm canopies as surrogates of wind farms. We also use the analytical flow solution over the hill, mainly to validate the simulations and to understand the dependence of AGW properties on the identified flow parameters, i.e. Froude number.    

The remaining part of this paper is structured as follows: \Cref{sec:flow_sce} gives the flow scenarios studied in this research. In \Cref{sec:methodology}, the methodology used to carry out this study and the methods used to analyze the LES data are described. The numerical models and setups for the hill and wind farm canopy cases are also explained in this section. \Cref{sec:results_woa} presents the results, discussed under various sub-sections based on addressing the dependence of reflections on parameters identified critical to model IGWs. Likewise, \Cref{sec:results_wc} gives the results tested with the wind farm canopy.  In \Cref{sec:conclu}, the conclusions of this research are presented as initial recommendations to set up simulations involving atmospheric gravity waves.  

\section{Flow Scenarios}
\label{sec:flow_sce}
This study is mainly based on flow over an isolated bell-shaped hill, which is used in various meteorology and earth sciences studies, such as \citet{Vosper2020} and \citet{snyder_thompson_eskridge_lawson_castro_lee_hunt_ogawa_1985}. The significant attractions of initially considering a hill instead of a wind farm include computational affordability, the availability of the analytical solution for validation, and simplicity. Moreover, the primary aim of this study is to handle atmospheric gravity waves in numerical simulations. Thus, the wave source is less important. Still, the simulated hill heights are analogous to typical wind turbine rotor tip heights, and the hill half-length to wind farm length. In the second flow case, we study flow through wind farm canopies to extend the findings from the hill case to wind farm applications. Unlike the rigid hill, a wind farm canopy is a wind farm-sized porous region that includes the drag force of the entire wind farm to the flow. Both flow scenarios are discussed in the following subsections to give a clear picture of this study.

\subsection{Flow over Hill}
\label{subsec:flow_woa}
Flow over the 2D bell-shaped WOA hill is solved both analytically and numerically. In both cases, the WOA hill is a surface with changing height ($h$) along the hill length ($l$). Thus, length and height are the only variables critical regarding the hill shape, defined by the following equation. 
\begin{gather}
    h(x) = \frac{h_{max}}{1+(x/l)^2}
\end{gather}
 
The hill height $h(x)$ varies with its length, with the maximum being $h_{max}$. Conditions like potential flow were adopted to keep the solution linear to compare the LES with the linear theory of gravity waves. To this end, surface roughness and Coriolis were excluded, and only a linearly stratified free atmosphere was solved with uniform inflow. Although, the turbulence sources are excluded from the setup, but the LES turbulence model is still enabled should there be any turbulence due to flow separation behind the hill. 

The computationally inexpensive semi-analytical flow solution over the hill was mainly used to understand the properties of IGWs and validate the numerical solutions. Besides, we use the analytical solution to develop the R-RMSE metric to analyze the simulation data, which will be discussed in \cref{subsec:quant_refl}. The following subsection gives an overview of this semi-analytical solution and how it is used in this study.   
  
\subsubsection{Semi-Analytical Solution}
\label{subsubsec:semi_analy}
The linear wave theory, particularly the Taylor-Goldstein equations, is commonly used to study atmospheric gravity waves in meteorology.  \citet{Allaerts_LBoW_-_Linear_2022} used these equations to develop a Python module called Linear Buoyancy Wave Package (LBoW) to solve linear buoyancy wave problems. In this research, we use a part of this code that computes a semi-analytical steady-state solution of the uniform, stratified flow over the WOA hill. This code relies on a grid to solve the linear theory, as the solution is performed in the frequency coordinates. Fast Fourier Transform(FFT) is done numerically in this solution, which makes the approach semi-analytical. The solution is independent of the grid size in the vertical direction but not in the horizontal. A solution at any vertical level can be acquired without requiring a prior solution at lower and higher levels. The grid dependence in the horizontal direction is linked to the FFT accuracy. FFT deviates from the theoretical Fourier transform of the bell-shaped hill for high horizontal wave numbers because of round-off errors. Mesh size in the range of $20~m$ to $100~m$ is recommended to compute the semi-analytical solution.

\begin{figure}[h]
    \centering
    \includegraphics[width=1\textwidth,trim={0cm 0cm 0 0cm},clip]{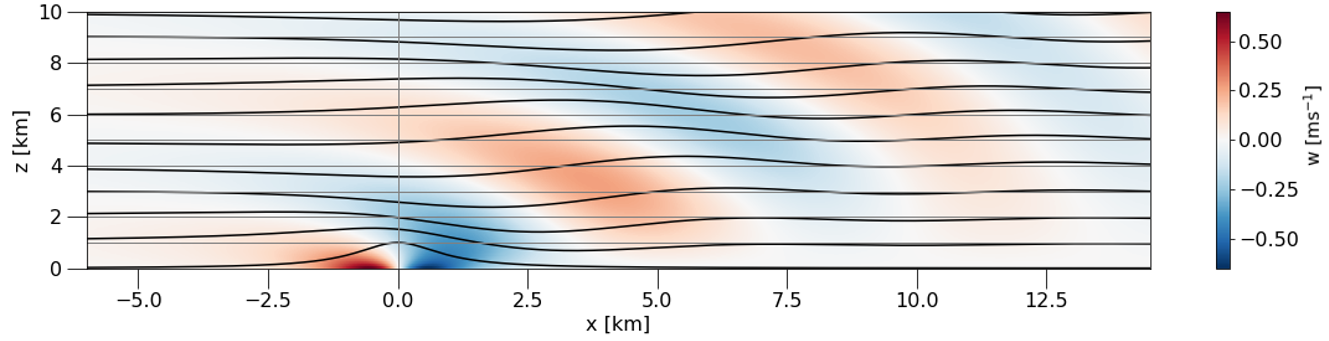}
    \caption{Steady-state analytical solution of flow over the hill at $x = 0 km$. The color contours represent vertical velocity and the streamlines show the propagation of the wave. \citet{Allaerts_LBoW_-_Linear_2022}}
    \label{fig:anly_woa}
\end{figure}

The steady-state solution of uniform flow over the hill under conditions like the stable free atmosphere is given in  \cref{fig:anly_woa}. The hill is placed directly in the linearly stratified atmosphere, and the upwards flow deflection by the hill triggers only the IGWs. The vertical velocity contour and streamlines clearly show the propagation of gravity waves in the vertical direction. Depending on flow conditions, the waves can be propagating or evanescent; the energy dissipates with height for the latter. It is essential to realize that the wave seen in \cref{fig:anly_woa} is a wave spectrum, where the wave properties depend mainly on the free-atmosphere Froude number and the hill shape and size. Various wavelengths are triggered depending on the flow's interaction with the hill. The wavelength with the highest amplitude is the dominant wave, which depends on the hill length. However, the apparent wave in the solution results from all superimposed waves, and the apparent wavelength differs from the dominant wavelength. Thus, the apparent wavelength is termed the effective wavelength. 

\begin{figure}
    \includegraphics[width=1\textwidth,trim={0cm 0.5cm 0 0.5cm},clip]{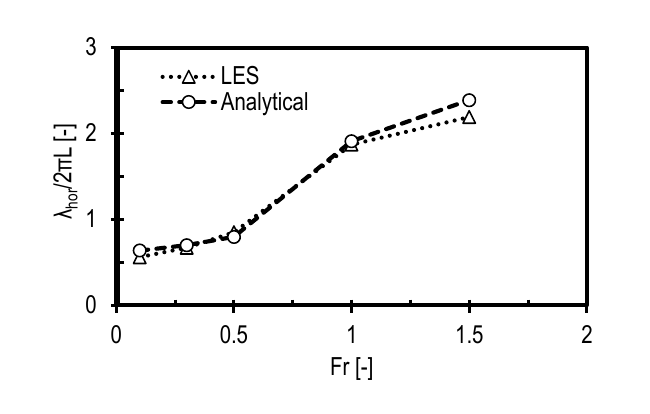}
    \caption{The variation of effective horizontal wavelength with Froude number, computed from the vertical velocity at mid-height of the domain for both LES and analytical solution for flow over the WOA hill.}
    \label{fig:hor_fr}
\end{figure}

The properties of IGWs depend on the variables like velocity ($U$), Brunt-V\"ais\"all\"a frequency ($N = \sqrt{(g/\theta)(\odv{\theta}{z})}$), hill half-length ($L$), and hill maximum height ($H$). Employing similarity theory, we normalize these variables to get two physical parameters, the Froude Number (Fr) and the Slope Parameter ($S_h$), where $Fr$ = $U/NL$ and $S_h$ = $H/L$.  We notice from the analytical solution that the wave amplitude depends more on the hill slope, particularly the height, whereas the wavelengths and direction of IGWs depend on $Fr$. The effective horizontal wavelength: ($\lambda_{hor}$) depends more on the hill length for low $Fr$, but for $Fr > 0.5$, it depends more on the scorer parameter ($N/U$), which is the reciprocal of the buoyancy length. This can be seen in \cref{fig:hor_fr}, where the normalized $\lambda_{hor}$, calculated from the semi-analytical solution and LES, is plotted against $Fr$. $\lambda_{hor}$ is twice the distance between the global maxima and minima on a vertical velocity plot at a height along the domain length. $\lambda_{hor}$ is dictated by the wave spectrum and the critical wave number, the scorer parameter, where the transition from propagating to the evanescent regime happens. The dominant wavelength corresponds to $1/L$ on the spectrum and is close to the effective wavelength when the critical wave number is greater than $1/L$ or for $Fr > 0.5$. In this case, the natural spectrum is preserved and propagating. On the other hand, for $Fr < 0.5 $, the critical wave number is smaller than $1/L$; thus, the spectrum becomes evanescent, and the dominant wave is dissipated. Therefore, $\lambda_{hor}$ depends more on the length than the scorer parameter. The effective vertical wavelength ($\lambda_{ver}$) depends only on the scorer parameter, which is proportional to the vertical wave number. Altogether, the effective wavelengths can be greater than the prominent length scales, i.e., the hill length and height; thus, domains scaled with the hill length and height might be inappropriate.

This study focuses only on the Froude number as it relates to gravity waves' properties critical to the simulation setup. The Slope Parameter is less critical for this research as we are not interested in flow scenarios created by varying it. Thus, the slope of the hill ($S_h$) is kept the same. Otherwise, the amplitude and wave spectrum would change, and comparisons of simulations with different hill shapes would be inconsistent. Time is another physical parameter not considered in this research; we keep it non-dimensionally constant. The only possible variation beyond the steady state is the build-up of reflections due to energy accumulation. Even the accumulation of the strongest reflections is not abrupt, as energy travels with group velocity and depends on advection speed, which is much longer than the buoyancy period for practical wind farm flow conditions.

\subsection{Flow through Wind Farm Canopy}
\label{subsec:flow_wc}
Wind farm flow interactions with the atmosphere involve a wide range of length scales. When focusing on only wind farm-induced IGWs, large length scales are critical as the expected IGW wavelengths are several kilometers; on the scale of the wind farm length. Since the intra-farm interactions is not the focus of this study, wind farm canopies are a convenient way to model wind farms without the complexity of modelling individual wind turbines.  The concept of wind farm canopy was introduced by \citet{markfort} through an analytical model to represent large wind farms in periodic weather prediction models. In our work, we use a similar approach to include the cumulative drag force of a wind farm of a given size and a kind of wind turbine on the flow. The total thrust distributed over the volume of a wind farm is calculated based on a fixed wind farm thrust coefficient ($C_{ft} = \frac{\pi C_t}{4 S_x S_y}$), and the velocity, which is sourced from cells centers within the wind farm canopy. The numerical implementation of the wind farm canopy is described in \cref{subsec:num_setup}. The wind turbine drag force per unit volume, which covers the turbine distancing, is given as,
\begin{equation}
\label{eq:fd_per_v}
    \frac{F_D}{V} = \frac{1}{2} \rho U^2 \frac{\pi C_t}{4 S_x S_y} \frac{1}{H}
\end{equation}

Where $H$ is the canopy top height, and $S_x$ and $S_y$ are non-dimensional lateral and streamwise turbine distancing, normalized with the rotor diameter, roughly accounting for the number of turbines. $C_{ft}$ is $C_t$ equivalent for an entire wind farm; the thrust coefficient of a whole wind farm. Multiplying \cref{eq:fd_per_v} with the total volume of a wind farm gives the entire wind farm's drag force.

In this second flow scenario, we simulate flow through wind farm canopies to test the findings from the hill case for wind farms. This is important because the flow deflected by a wind farm can be significantly less than that of a hill, and the wave properties may look different for the same flow conditions. Therefore, the flow conditions considered for wind farm canopies are the same as those for the hill case, except that the hill is replaced by a porous region that imposes the wind farm drag on the flow.

\section{Methodology}
\label{sec:methodology}
\subsection{Simulation Parameters and Setup}
\label{subsec:num_setup}
A set of non-dimensional parameters govern the flow over terrain and through wind farms under linearly stratified atmospheric conditions. These parameters can be acquired by normalizing the flow equations or performing dimensional analysis of the variables in this study that are given in \cref{tab:imporvara} along with their practical values in wind applications. The only physical parameters are Froude Number and Slope Parameter, already introduced in \cref{subsubsec:semi_analy}. The other four are simulation parameters, including $\tilde{X}$ (the domain length normalized with $\lambda_{hor}$), $\tilde{L}_z$, and $\tilde{L}_d$, (the domain height and damping thickness normalized with $\lambda_{ver}$). Finally, $\xi$, the damping coefficient normalized with Brunt-Vaisala frequency. Grid structure, resolution, and time step could be important simulation parameters for gravity wave activity for very low $Fr$. Because the flow interactions can trigger sub-grid scale wavelengths, the cell faces may act like boundaries for those waves. Likewise, the frequencies of some waves in the spectrum could be shorter than the simulation time steps, leading to an unresolved fraction of the spectrum. However, $Fr$ for wind farm applications would most likely be between $0.1$ to $0.5$. we did a grid independence study for this range of $Fr$. We found that grid resolutions used in wind farm LES are appropriate for modeling wind-farm-induced AGWs. A thorough investigation for very low $Fr$ falls more on the sensitivity part of the numerical modeling, which is not in the scope of this research. 

\begin{table}
    \centering
    \caption{Main variables involved in simulating atmospheric gravity waves under linearly stratified free atmospheric conditions.}
    \begin{tabular}{lll}
        \toprule
        {Variable} & {Variable Type} & {Range of Tested Values} \\
        \midrule
        Velocity ($U$) & Flow & $1~ms^{-1}$ to $1~ms^{-1}$  \\
        Brunt-Vaisala frequency ($N$) & Flow & $0.005~s^{-1}$ to $0.02~s^{-1}$   \\
        Half-Length of Hill or Canopy Length ($L$) & Shape & $1~km$1 to $15~km$\\
        Hill or Canopy Height ($H$) & Shape & $16~m$16 to $240~m$\\
        Domain Length ($X$) & Numerical & $0.5~km$ to $200~km$\\
        Domain Height ($L_z$) & Numerical & $0.3~km$ to $40~km$\\
        Damping thickness ($L_d$) & Damping characteristic & $0.3~km$ to $45~km$\\
        Damping coefficient ($1/\tau$) & Damping characteristic & $0.001~s^{-1}$ to $0.5~s^{-1}$\\
        \bottomrule
    \end{tabular}
    \label{tab:imporvara}
\end{table}


LES of flow over the WOA hill and through the wind farm canopy is carried out with Simulator for Wind Farm Applications (SOWFA) \citet{churchfield2012overview}. Based on OpenFOAM, this code is mainly used for the LES of atmospheric flows over terrain and through wind farms, where a one-equation model is commonly used for sub-grid-scale turbulence modelling. SOWFA has actuator models for wind turbine aerodynamics that can be coupled with aero/servo/elastic tools. Moreover, it can take inflow data from mesoscale weather data, and terrain can be included through non-conformal meshes. This study solves the incompressible Navier-Stokes equations under non-hydrostatic conditions and with the Boussinesq approximation for buoyancy.  Equations for continuity, momentum, and potential temperature are typically used in the LES of atmospheric flows. The description of solving these equations with SOWFA is given by \citet{Churchfield2012}. For simulations with the wind farm canopy, the drag force of the wind farm is added to the momentum equation as a body force. Rayleigh damping force is also a body force applied only in the forcing zones, included in the simulations through the momentum equation.

\begin{figure}[h]
    \includegraphics[width=1\textwidth]{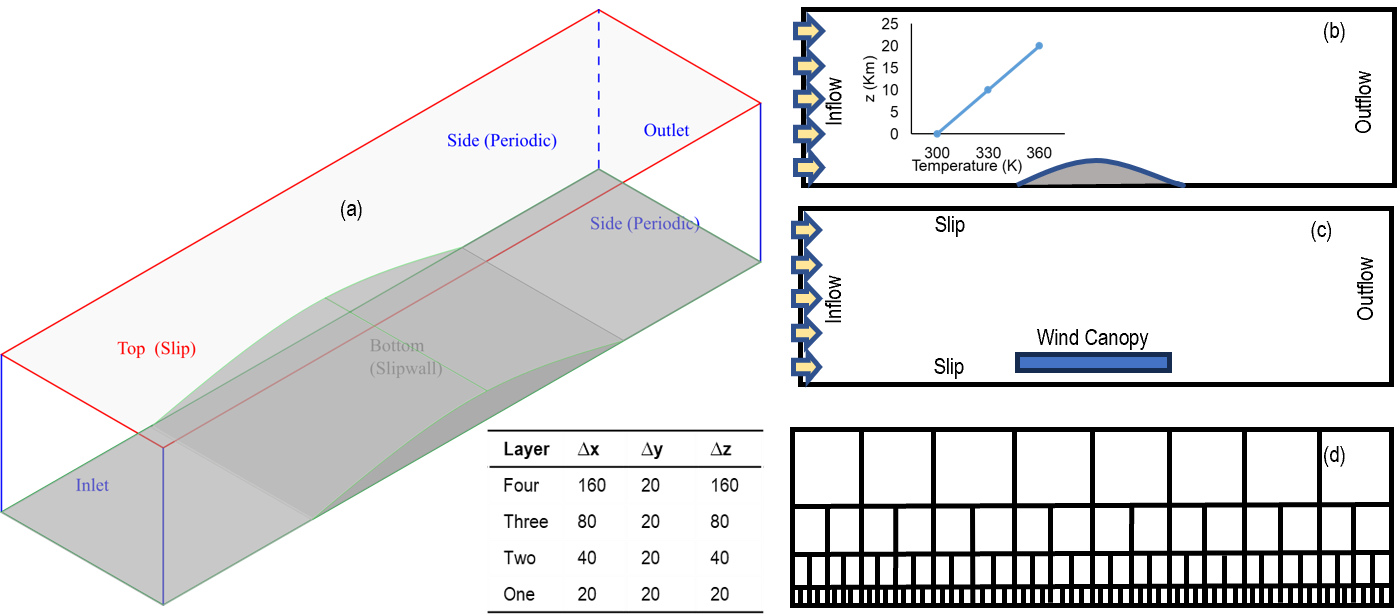}
    \caption{Numerical setup: (a) isometric view with a schematic hill and (b) side view with a hill, and (c) with wind farm canopy. (d) The grid topology alongside layer-wise grid size.}
    \label{fig:setup}
\end{figure}

\Cref{fig:setup} presents the numerical setup for the hill and wind farm canopy and the mesh size and topology. Unlike the commonly used periodic boundary conditions to model AGWs, the flow is driven in and out of the domain by inflow/outflow boundary conditions. Periodic boundary conditions are used in the transverse direction only. Both the ground and the top are modeled as slip boundary conditions. This also means there is no wind shear, and the domain is fed a uniform wind velocity at the inlet. The temperature distribution is linear in the vertical direction, with constant Brunt-V\"ais\"al\"a frequency. The ground has no heat flux. Thus, the conditions implemented are closest to a stable free atmosphere without the ABL and inversion layer.

A surface definition for the WOA hill is created with a Python script and the computational mesh conforms to the hill. A mesh with layered refinement is used. Resolution near the surface is fine, but it coarsens at higher altitudes. The cell sizes for most cases in each of the four layers of the domain are shown in \cref{fig:setup}. The domain for all cases is $100~m$ in the $y$, whereas $x$ and $z$ extents varied as fractions and multiples of the effective horizontal and vertical wavelengths, respectively. 

We implement wind farm canopies to test the findings from the WOA case for the wind farms. It is an ideal approach to reduce computational resources, which is desirable as we simulate hundreds of cases to test the reflection trends. \Cref{fig:setup}(c) shows the side view of the setup with a wind farm canopy. The numerical setup with the wind farm canopy is the same as that of the WOA hill, except that the hill is replaced with wind farm drag as a body force. Thrust per unit cell volume is calculated for each cell in the canopy by sourcing velocity from each cell. 

\subsection{Rayleigh Damping}
\label{subsec:ray_damp}
RDLs are zones adjacent to the reflective boundaries in the simulation domain and far from the region of interest, like a wind farm. The primary role of an RDL is to dissipate the energy brought into the zone by AGWs. Only the vertical velocity is damped unless mentioned because the prominent perturbation is the vertically deflected velocity by the hill or the wind farm canopy. In the literature, RDL is used at the top boundary only and a fringe region, used only with periodic boundary conditions, at the outlet, similar to an RDL in damping AGWS, but it also recycles the flow to the inlet. However, in a numerical setup with inflowOutflow boundary conditions, RDLs may be required at various boundaries. The RDL at the inlet should filter the incoming turbulence; damping incoming turbulence into the boundary layer is undesirable in wind farm simulations. However, we did not include inflow turbulence in this study. 

\subsection{Quantifying Reflections}
\label{subsec:quant_refl}
Reflectivity is quantified by the method proposed by \citet{Allaerts2017}, which is a modification of the procedure initially given by \citet{Taylor2007}. The reflection coefficient ($Cr$) is one of the two primary tools used here to analyze the simulation data, and its calculation can be summarised in the following two steps. First, the vertical plane's vertical velocity is decomposed into upward and downward moving waves. Then, the energy associated with these decomposed waves is calculated, and the reflection coefficient is estimated by taking the ratio of downwards to upward propagating energy. The $Cr$ metric sometimes gives inconsistent outputs, especially for low $Fr$ and $L$, possibly because the spectrum includes mainly high-frequency waves. Also, taking the ratio of energies could be misleading in some cases, as the reflected waves can be directed more upwards, thus reducing the $Cr$ value. Therefore, visual inspection of the simulation fields, mainly vertical velocity, is critical to ensure the value predicted is realistic; improving or replacing this metric is future work.   

Relative-Root Mean Square Error (R-RMSE) is the other metric used only for the hill case. It is calculated by comparing the numerical solution with the analytical solution.  R-RMSE is introduced to ensure a consistent scale for comparison of the simulations that involve several flow variables with a broad range of values. In this regard, the vertical velocity fields from the numerical and analytical solutions are normalized with the analytical solution, thus making it Relative-RMSE. The vertical velocity is always taken at $tN = 300$, which converts into $2.083$, $4.167$, or $8.33~hrs$ for $N = 0.02, ~0.01$ or $0.005~s^{-1}$, respectively. The normalized vertical velocities are then processed in the typical way of calculating RMS error. R-RMSE is sensitive to sample space size. R-RMSE for a large domain is lesser than the same case with a smaller domain. A way to circumvent this is to consider the same section of the domain, around the hill or canopy, but this may leave out the reflections building close to the boundaries. One of the reasons why we use R-RMSE is because $Cr$ can only quantify downward wave reflections, but it cannot be used to see whether there are spurious wave sources. Thus, we analyze the results using both metrics, and generally, R-RMSE detects the same patterns and trends shown on $Cr$ plots.  In all cases, unless mentioned, vertical velocity on a vertical plane at the mid-point of the domain is processed to calculate both $Cr$ and R-RMSE. To get a single value for $Cr$ and R-RMSE, the downwards and upwards propagating energies, and point-wise R-RMSE on the vertical plane are averaged in the x and z directions. We follow the criterion $Cr$ and R-RMSE $< 10\%$ for reasonably reflection-free simulations.

\subsection{Simulation Sets}
\label{subsec:sets}
Simulations were run initially to acquire a base case and then use it to explore reflection dependency on the non-dimensional parameters discussed previously. Various combinations of damping layers with different combinations of characteristics were simulated to investigate the most appropriate configuration with minimum reflections. The details on the configurations explored are given in the Appendix. Besides exploring the configuration of damping layers, three simulation sets are designed as given in \cref{table: sim-set}. The first set of simulations investigates the impact of damping characteristics on reflections and their optimal values for a range of $Fr$. Set 1(a) explores the trend of reflections for varying the damping coefficient only. The reflection trend was tested for a range of $Fr$ values, i.e., $0.1$ to $1.5$, where $Fr$ was adjusted by changing $U$ and $L$. In set 1(b), the damping coefficient and thickness were changed for $Fr$= $0.1$ and $0.5$, where $Fr$ was adjusted by changing either $N$ and $U$, or $U$, $N$, and $L$. Thus ensuring the validity of the optimal damping characteristics for dynamically similar solutions. 

The second simulation set inspects the impact of the domain length on the reflections. In this case, the domain length was varied for two $Fr$. Further, two hill lengths, $5~km$ and $10~km$, were simulated for each $Fr$. The third set investigates the domain heights' effect on the reflections. To this end, for a couple of $Fr$, only domain height was varied in proportion to the expected ($\lambda_\mathrm{ver}$).  Finally, these three simulation sets were reproduced for the wind farm canopy to test the validity of the findings from the hill case to wind farms. 

\begin{table}
  \caption{Simulation set to conduct the LES study for the hill case.}
  \begin{tabularx}{0.8\textwidth}{l *{5}{c}}
    \toprule
    Set & \multicolumn{2}{c}{Parameters to Investigate}
             & \multicolumn{2}{c}{Variables to Vary}
                        & Number of Simulations               \\
    \midrule
      &  Physical & Simulation &   Physical  &  Simulation        &    \\
    \midrule
    1(a)  &  Fr [0.1, 0.3, 0.5, 1.0, 1.5]  & $\xi$, $\tilde{L}_d$  &   U, L     &  $1/\tau$       &   25 \\
    1(b)  &  Fr [0.1, 0.5]                 & $\xi$, $\tilde{L}_d$  &   U, N, L  &  $1/\tau$, $L_d$ &   120 \\
    2     &  Fr [0.1, 0.5]                 & $\tilde{X}$           &   Fr, L [5, 10 km]    &  X   &   20 \\
    3     &  Fr [0.1, 0.5]                 & $\tilde{L}_z$         &   Fr       &  $L_z$       &   12 \\
    \bottomrule
  \end{tabularx}
\label{table: sim-set}
\end{table}

\section{Results (The Hill scenario)}
\label{sec:results_woa}
Configuration of damping layers, damping characteristics, domain length, and domain height are the four most important simulation aspects to tune in setting up simulations involving AGWs. Initially, the damping layers' location was considered to develop a base case for flow over the WOA hill. To this end, various configurations were simulated to determine the arrangement of damping layers that give minimum reflections. The base case is a setup with reflections less than 10\%, and any simulation fulfilling this criterion is considered reasonably reflection-free in this study. A brief investigation was carried out to determine the positioning of the damping layers; RDLs at the inlet, outlet, and top boundaries give minimum reflections. Thus, all simulations used this configuration of the RDLs.  The remaining three aspects were thoroughly investigated to provide a reference to determine the damping characteristics and domain size in simulating flow conditions practical to wind farm applications.

\subsection{Dynamic Similarity and Quantification Metrics}
\label{subsec:dynamic}
\begin{figure}
    \hfill
    \subcaptionbox{Reflection coefficient as a function of $\xi$ \label{fig:dyn_fr0.1}}{\includegraphics[width=8.8cm,trim={0cm 0.5cm 0 0.5cm},clip]{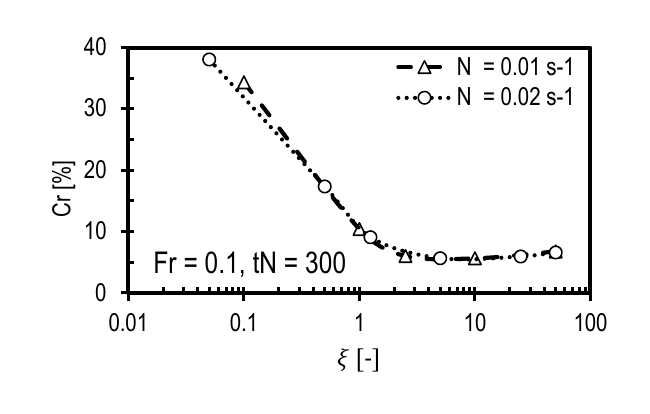}}
    \hfill
    \subcaptionbox{Relative-root mean square error as a function of $\xi$ \label{fig:dyn_fr0.5}}{\includegraphics[width=8.8cm,trim={0cm 0.5cm 0 0.5cm},clip]{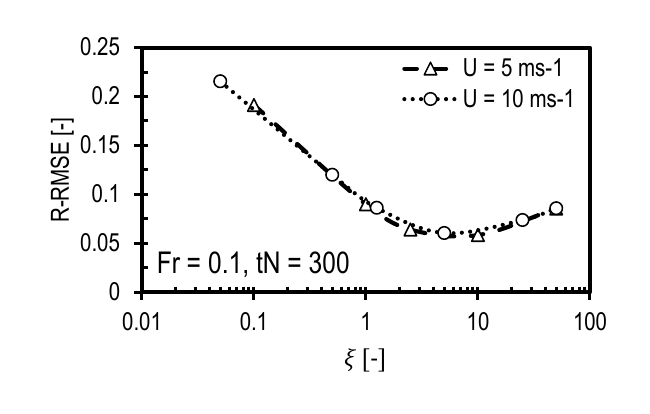}}
    \hfill\strut
    \caption{Dynamically similar cases for Fr = 0.1 by varying damping coefficient and: (a) the buoyancy time scale shown in terms of $Cr$, (b) the length scale shown in terms of R-RMSE.}
    \label{fig:dynamic}
\end{figure}

We started the analysis by testing whether the non-dimensional parameters discussed in \cref{subsec:num_setup} are sufficient to ensure dynamic similarity. The simulations performed had a setup with a damping layer thickness and domain height $1.5$ times the effective vertical wavelength and a domain length five times the expected effective horizontal wavelength. The overlapping plots in \cref{fig:dynamic}, suggest the same solutions for varying the time and length scales.  In \cref{fig:dyn_fr0.1}, the buoyancy time was varied by changing the Brunt V\"ais\"al\"a frequency, but the Froude number was kept constant by adjusting only the hill length.  Another finding in this figure is the appropriateness of scaling the damping coefficient with the buoyancy time scale. While varying $N$, the damping coefficient was also varied to set the desired values of $\xi$. The length scale was varied by adjusting the inflow velocity and hill length to maintain $Fr = 0.1$; the R-RMSE against $\xi$ for this case is given in \cref{fig:dyn_fr0.5}.  Similar overlapping plots were seen for $Fr = 0.5$, which are not shown here to avoid redundancy.  \Cref{fig:dynamic} also shows the consistency in depicting the reflections by the $Cr$ and R-RMSE metrics; both suggest the suitable $\xi$ range for  $Fr~0.1$ to be $1$ to $10$. Therefore, we present the analysis here on wards using $Cr$ only.          

\subsection{Froude Number and Optimal Damping Coefficient}
\label{subsec:Fr}
\begin{figure}[H]
    \includegraphics[width=1.0\textwidth,trim={0cm 0.1cm 0 0.1cm},clip]{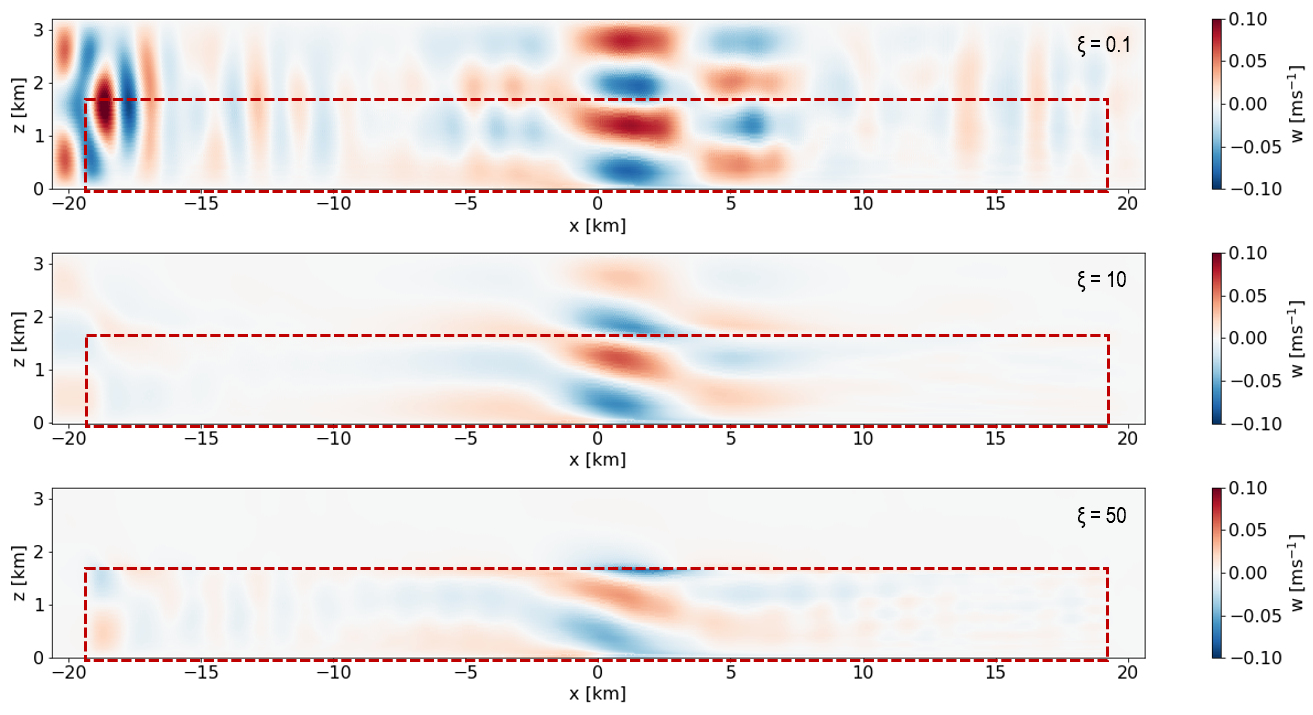}
    \caption{Contours of vertical velocity in a streamwise-oriented vertical plane at $tN~300$ for $Fr = 0.1$ with $\xi = 0.1$ (top), $\xi = 10$ (middle), and   $\xi = 50$ (bottom). The red box shows the non-damped domain, and everything outside is the RDL.}
    \label{fig:vel_cont}
\end{figure}

The optimal damping coefficient for various $Fr$ might differ, but the reflection pattern for varying $\xi$ is the same. Generally, very low damping coefficients lead to the highest reflections. $\xi > 10$ also enhance the reflections because the IGWs reflect off the damping layer instead of the boundaries. Strong reflections and energy accumulation can be seen visually in \cref{fig:vel_cont}(top) for $\xi = 0.1$, where they appear parallel to the inlet on the left and gradually contaminate the solution in the entire domain. \Cref{fig:vel_cont}(middle) shows vertical velocity contour for $\xi = 10$ that is least polluted by reflections and eneegy accumulation at the inlet. Moreover, the gradual IGW decay with height inside the top RDL is the evidence of suitable damping characteristics. Whereas, \cref{fig:vel_cont}(bottom) shows a vertical velocity contour for for $\xi = 50$, where IGWs are abruptly attenuated right at the start of the top damping layer. Also the accumulated waves appear at the end of inflow RDL as if it is a boundary. 

\begin{figure}[h]
    \includegraphics[width=0.8\textwidth,trim={0cm 0.5cm 0 0.5cm},clip]{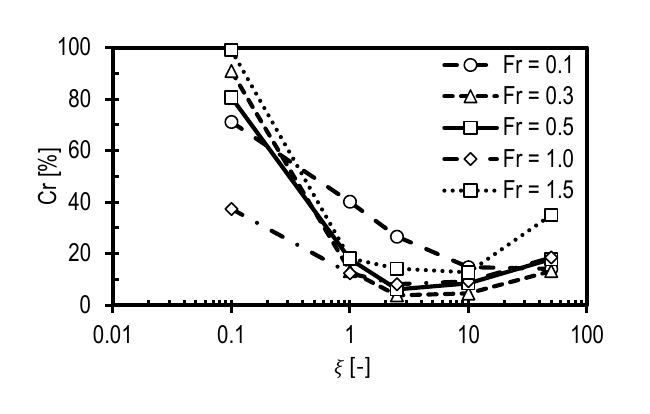}
    \caption{Reflection coefficient for a range of Froude numbers as a function of $\xi$  only.}
    \label{fig:Cr_vs_fr}
\end{figure}

We have already seen in \cref{subsec:dynamic} that the optimal damping coefficient for $Fr = 0.1$ is around $10$. However, \cref{fig:Cr_vs_fr} shows that for $Fr > 0.1$, the optimal damping coefficient could be around $2.5$. The inappropriateness of weak and very strong damping coefficients is also confirmed in this plot. The reflections are the highest for all cases with $\xi = 0.1$, and $\xi = 50$ is also less effective. For $Fr = 0.1$ and $1.5$, $Cr$ is not less than $10\%$, because the damping layer thickness is just one vertical wavelength; probably not sufficient for very low and supercritical $Fr$. These simulations had damping layer thickness and domain height equal to the effective vertical wavelength, and the domain length was two times the effective horizontal wavelength. Supercritical $Fr$ are less likely to occur for large-wind farms, whereas low $Fr$ values are very likely. Therefore, the following sections will thoroughly investigate $Fr = 0.1$ and $0.5$.  

\subsection{Impact of Damping Layer Thickness on Reflections}
\label{subsec:Ld}
So far, we have considered the impact of $\xi$ on the reflections while $L_d$ is equal to or greater than the vertical wavelength. However, the impact of damping characteristics on reflections is coupled. A weak, thick damping layer may have the same impact as a strong, thin layer. Therefore, determining the coupled impact of the damping characteristics and the minimum damping thickness is desirable, as knowing the minimum effective damping layer thickness can help reduce the computational load. 

\begin{figure}
    \hfill
    \subcaptionbox{Reflection coefficient as a function of damping characteristics for $Fr = 0.1$\label{fig:cr_vs_ld_fr0.1}}{\includegraphics[width=8.8cm,trim={0cm 0.5cm 0 0.5cm},clip]{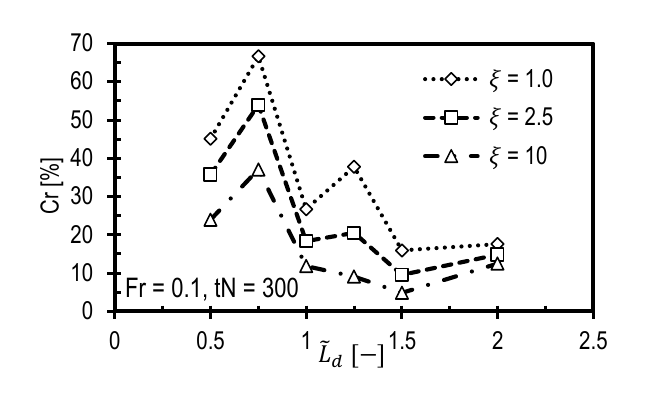}}
    \hfill
    \subcaptionbox{Reflection coefficient as a function of damping characteristics for $Fr = 0.5$\label{fig:cr_vs_ld_fr0.5}}{\includegraphics[width=8.8cm,trim={0cm 0.5cm 0 0.5cm},clip]{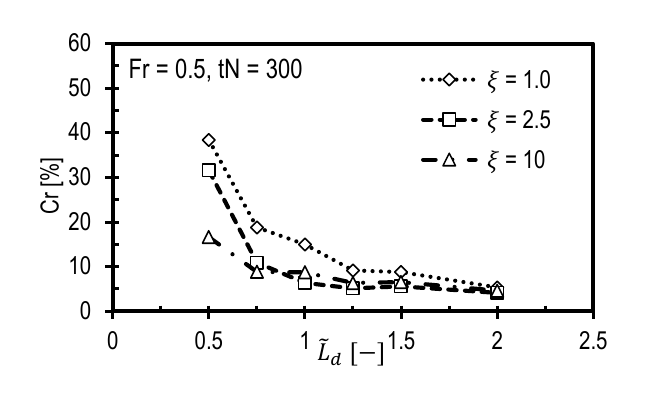}}
    \hfill\strut
    \caption{The coupled impact of damping characteristics on reflections for: (a) $Fr = 0.1$, and (b) $Fr = 0.5$.}
    \label{fig:Ld}
\end{figure}

The coupled impact of the damping characteristics is investigated for $Fr = 0.1$ and $Fr = 0.5$, where the damping thickness and coefficient are varied simultaneously. \Cref{fig:Ld} shows that the thicker damping layers for all damping coefficients would reduce the reflections. Also, a damping layer accommodating one effective vertical wavelength seems effective with an optimal damping coefficient. As shown in \cref{fig:cr_vs_ld_fr0.1}, $Cr$ is minimum for all $\tilde{L}_d$ when $\xi$ is $10$. However, optimizing the damping characteristics and the setup for low $Fr$ is more challenging. The abrupt variations in reflection for varying damping layer thickness indicate one of these challenges. Since wavelengths become shorter for low $Fr$, the wave spectrum could be more sensitive to length scaling. Moreover, the inclination of the waves varies with $Fr$ and becomes more aligned to the horizontal for low $Fr$, complicating the interaction with the background advecting flow. This also hints at the RDLs lacking to eliminate energy accumulation at the inlet and only do enough to delay its propagating back into the domain. We believe the energy accumulation is higher for low $Fr$ as the wave speed is faster than advection and the wave fronts fall more directly onto the inlet than that of a higher $Fr$ case. Thus, domain heights capturing wave fronts partially have more energy accumulation, and in turn more contamination.   

\Cref{fig:cr_vs_ld_fr0.5} indicates the same findings for $Fr = 0.5$. We notice that $\xi = 2.5$ is optimal with $\tilde{L}_d > 1.0$, and the impact of $\xi = 10$ is also not very different and is more effective for $\tilde{L}_d < 1.0$. The damping layer thickness required with $Fr = 0.1$ is slightly bigger than that of $Fr = 0.5$ to limit the reflections to the same levels. This can be established by comparing the optimal setups for $Fr = 0.1$ and $Fr = 0.5$, where the reflections are limited to $5\%$ when $\tilde{L}_d$ is $1.2$ for $Fr = 0.5$ and $\tilde{L}_d$ is $1.5$ for $Fr = 0.1$.  In summary, the damping layer thickness should be greater than one effective vertical wavelength to dampen the IGWs. This aligns with the recommendation of \citep{klemp-Lilly1978} but considers the entire wave spectrum instead of their approach of analyzing individual wave numbers.        

\subsection{Domain Length Impact on Reflections}
\label{subsec:DL}
Intuitively, the location of the top boundary is more important than positioning other boundaries, as the gravity waves travel upwards, and reflections are mainly expected from the top boundary. However, the accumulation of energy at the inlet boundary, shown in \cref{fig:ref_vc} and \cref{fig:vel_cont}, indicates the importance of appropriately positioning the inlet and outlet relative to the hill or wind farm. Placing boundaries far from the zone of interest may be costly for LES studies. Also, simulations can be run for a limited time to prevent the reflections from reaching the zone of interest. Which may be undesirable, especially, as long diurnal simulations become more common. Moreover, the contamination upstream of a wind farm may affect the flow, and unrealistic conditions develop. Therefore, knowing the shortest possible domain length is important to save computational resources and time. To this end, a set of simulations, given in \cref{table: sim-set} as set 2, were performed with two hill lengths for $Fr = 0.1$ and $Fr = 0.5$, and the domain length was varied in the range of $0.5$ to $ 4.0\lambda_{hor}$. Instead of simulating for several damping characteristics, we used the results from \cref{subsec:Ld} with $\xi = 10$ for $Fr = 0.1$ and $\xi = 2.5$ for $Fr = 0.5$.  $\tilde{L}_d$ was set to $1.5$ for $Fr = 0.1$ and $1.2$ for $Fr = 0.5$.

\begin{figure}[h]
    \centering
    \hfill
    \subcaptionbox{Reflection coefficient as a function of domain length\label{fig:x_vs_l}}{\includegraphics[width=8.8cm,trim={0cm 0.5cm 0 0.5cm},clip]{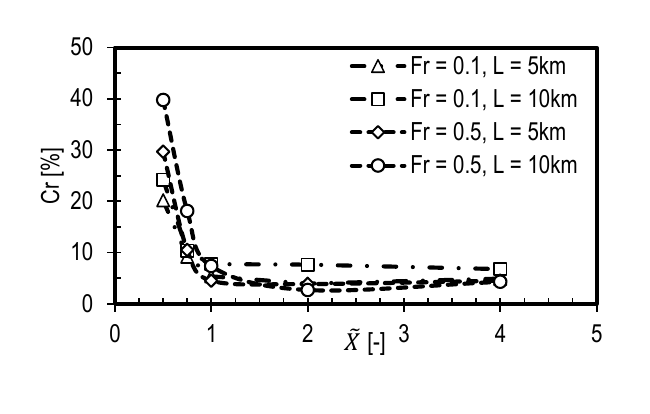}}
    \hfill
    \subcaptionbox{Reflection coefficient as a function of domain height\label{fig:cr_vs_lz}}{\includegraphics[width=8.8cm,trim={0cm 0.5cm 0 0.5cm},clip]{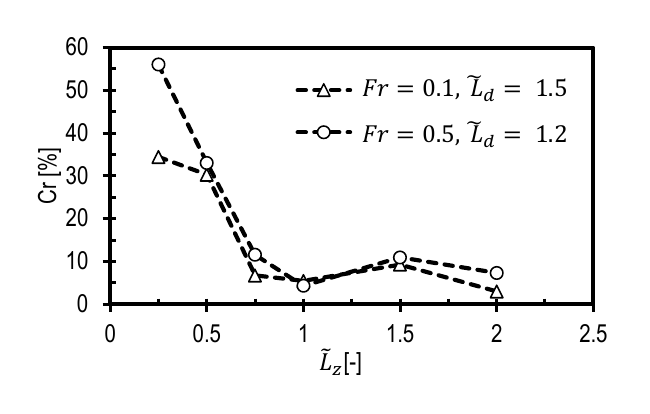}}
    \hfill\strut
    \caption{The reflection impact of (a) domain length for $Fr = 0.1$ and $Fr = 0.5$ with hill lengths $5$ and $10~km$, and (b) domain height for $Fr = 0.1$ and $Fr = 0.5$.}
    \label{fig:domainlength}
\end{figure}
  
\Cref{fig:x_vs_l} shows the impact of domain length on reflections. The reflections show the same trend for $Fr = 0.1$ and $Fr = 0.5$ with $L = 5$ and $10~km$. The domain length should be at least one effective horizontal wavelength, as $Cr$ increases abruptly for shorter domains. We emphasize the discussion in \cref{subsec:flow_woa} about the variation in effective horizontal wavelength with $Fr$. We established that $\lambda_{hor}$ depends on $L$ and $U/N$ for $Fr < 0.5$ and much more on $U/N$ for $Fr > 0.5$. In other words, the domain length should not be scaled with the $L$. Instead, calculate the expected effective horizontal wavelength from the linear theory and set the domain length to accommodate at least one effective horizontal wave. Domain lengths over $\lambda_{hor}$ could be redundant and costly for short runs, a few hours if sufficient flow-through times can be met. But simulating the diurnal cycle may require domains several $\lambda_{hor}$ long, especially the inlet should be sufficiently far away. This prevents the accumulating waves at the inlet from reaching the wind farm. It is important to note that the simulated domains are symmetric about the hill apex; the distance between the inlet and outlet is the same from the center. In wind farm simulations, there could be a limit on the distance upstream of the wind farm to allow the flow to adjust with the pressure field created by the AGWs and to avoid artificial blockage \citep{Lanzilao2022}. This limit on domain length will be investigated with the complete atmospheric temperature structure in an upcoming study.

\subsection{Domain Height Impact on Reflection}
\label{subsec:Lz}
We anticipate that the height of the domain should scale to the effective vertical wavelength. To understand this, we designed a set of simulations, set 3 in \cref{table: sim-set}, by changing only the height of the domain in proportion to the expected $\lambda_{ver}$. Six domain heights in the range of ($\tilde{L}_z = L_z/\lambda_{ver} = 0.25$ to $2.0$) were simulated each for $Fr = 0.1$ and $0.5$. The domain length was equal to $\lambda_{hor}$ for all simulations, and the damping thickness was set to $1.5\lambda_{ver}$ for $Fr = 0.1$ and $1.2\lambda_{ver}$ for $Fr = 0.5$. Further, $\xi$ was set to $10$ for $Fr = 0.1$ and $2.5$ for $Fr = 0.5$.

\Cref{fig:cr_vs_lz} shows the reflection coefficient for varying the non-damped domain height. The major observation in this plot is the huge reduction in reflections for domain heights approaching $\lambda_{ver}$ from the lower side. As discussed in \cref{subsec:Ld}, the reflections appear more irregular for $Fr~0.1$ than $Fr~0.5$. There is a slight increase in reflections for $\tilde{L}_z$ $1.5$ for both $Fr$, but further increasing the domain height seems to reduce the reflections. Our experience with this study shows that taller domains allow more waves to reach the inlet than shorter domains. Since the waves are inclined, the wavefronts hit the top damping layer for short domains before reaching the inlet. However, if the domain is taller than one $\lambda_{ver}$, depending on $Fr$, the wavefronts farther from the source reach the inlet before hitting the top damping layer. Thus, the reflection pattern in \cref{fig:cr_vs_lz} is slightly irregular. In any case, it is important to keep the non-damped domain height around one effective vertical wavelength. This recommendation may change upon including the ABL and inversion layer in the simulations. The inversion height might be critical to setting the non-damped domain height.

\section{Results (Wind Farm Canopy)}
\label{sec:results_wc}
The investigation of flow over the hill provided a baseline for simulating AGWs under linearly stratified free atmospheric conditions. The next step in this research is to extend these findings to wind farm applications. Thus, the tests performed on the hill case are reproduced on the porous wind farm canopy. The simulation setup for each type of study is the same as the hill cases, except that the hill is replaced with a wind farm canopy. The canopy length and height correspond to the hill's half-length and maximum height, respectively. 

\subsection{Damping Characteristic Impact on Reflections}
\label{subsec:wc_fr_damp}

\begin{figure}[h]
    \centering
    \hfill
    \subcaptionbox{Reflection coefficient as a function of damping characteristics for $Fr~0.1$\label{wc_cr_vs_ld_fr0.1}}{\includegraphics[width=8.8cm,trim={0cm 0.5cm 0 0.5cm},clip]{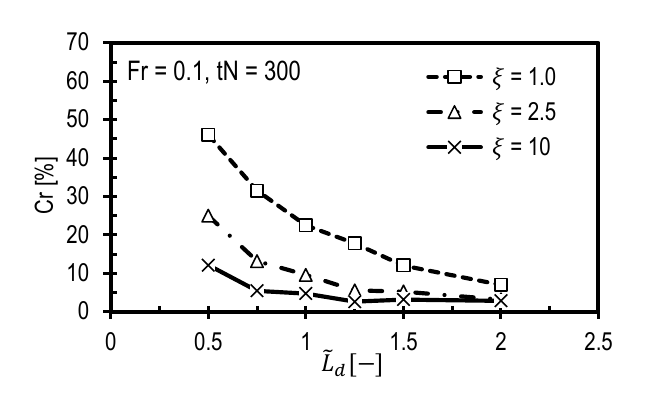}}
    \hfill
    \subcaptionbox{Reflection coefficient as a function of damping characteristics for $Fr~0.5$\label{fig:wc_cr_vs_ld_fr0.5}}{\includegraphics[width=8.8cm,trim={0cm 0.5cm 0 0.5cm},clip]{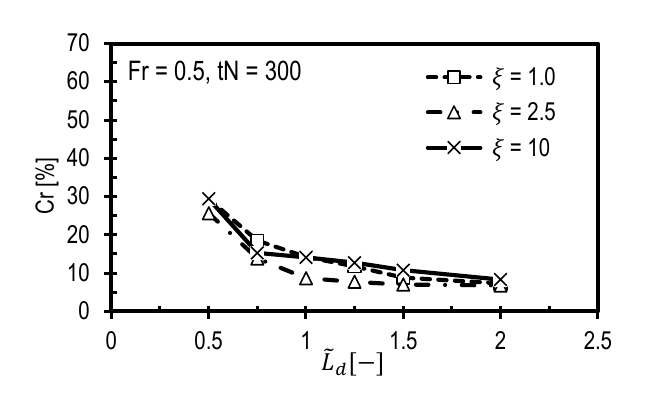}}
    \hfill\strut
    \caption{The impact of damping characteristics on reflections of wind farm canopy-induced IGWs.}
    \label{fig:wc_fr_damp}
\end{figure}

 The findings from the WOA case are almost equally applicable to the wind farm canopy setups. This can be seen in \cref{fig:wc_fr_damp}, plotting the coupled impact of damping characteristics on reflections. These plots are for $Fr = 0.1$ and $0.5$ at a non-dimensional time ($tN = 300$), which translates into dimensional time of either $4.12$ or $8.3$ hours, for $N = 0.02$ and $0.01~s^{-1}$, respectively. \Cref{fig:wc_fr_damp} shows similar variations in $Cr$ for the WFC to those seen in \cref{fig:Ld} for the hill, expect that $Cr$ variation for $Fr = 0.1$ for WFC is monotonic. This suggests a different wave spectrum for the WFC than that of the hill, both simulated under the same conditions. This aspect is addressed with visual plots in the next subsection. 
 
 Generally, thicker damping layers compensate for weaker damping coefficients, which can be noticed in \cref{fig:wc_fr_damp}. More importantly, for almost all $\xi$ and $\tilde{L}_d < 1$, $Cr$ exceeds the threshold we selected in this study. The suitable damping coefficient range for $Fr = 0.1$ is still $1.0$ to $10$, with $10$ being optimal for all damping layer thicknesses. Likewise, $\xi = 2.5$ is the optimal damping coefficient for $Fr = 0.5$, given in \cref{fig:wc_cr_vs_ld_fr0.5}, across all damping layer thicknesses. Further, $\xi = 1$ and $10$ appear slightly less effective than $\xi = 2.5$.

\subsection{Domain Size Impact on Reflections}
\label{subsec:wc_Lz}
The domain size is critical in wind farm simulations to capture various phenomena in and around the wind farm. The contribution of wind farm-induced AGWs to the global blockage effect requires including the temperature structure of the atmosphere to heights that admit at least one IGW wavelength. This could mean simulating domains taller than conventional ones that include only the ABL. Based on the findings about the domain size from the hill study discussed earlier in this paper, the domains to simulate the wind farms may be even larger than what could be required to capture the IGWs. On the other hand, it is desirable to simulate as small domains as possible to reduce the computational cost. Therefore, investigating the limits on domain size while simulating wind farm-induced AGWs is essential. To extend the hill results, we reproduce simulation sets $2$ and $3$ from \cref{table: sim-set} with the wind farm canopy. The setups are the same as that of the hill cases described in \cref{subsec:DL} and \cref{subsec:Lz}, except that the hill is replaced with a wind farm canopy in each simulation. For WFC lengths $10~km$, the WFC starts at $40~m$ and goes to $160~m$ in vertical. Therefore, the WFC thrust is stronger for the $L = 10km$ cases.

\begin{figure}[h]
    \includegraphics[width=1.0\textwidth,trim={0cm 2cm 0 2cm},clip]{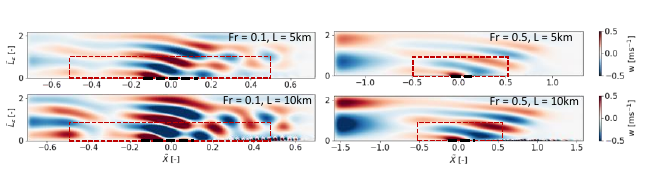}
    \caption{Vertical velocity of flow through the wind farm canopy with lengths $L = 5$ and $10~km$ for $Fr = 0.1$ and $0.5$, and $\tilde{X} = 1.0$. The black dashed line represents the canopy length, and the region inside the dashed red box is the non-damped domain.}
    \label{fig:wContours_canopy}
\end{figure}

\begin{figure}[h]
    \centering
    \hfill
    \subcaptionbox{Reflection coefficient as a function of domain length\label{fig:wc_x}}{\includegraphics[width=8.8cm,trim={0cm 0.5cm 0 0.5cm},clip]{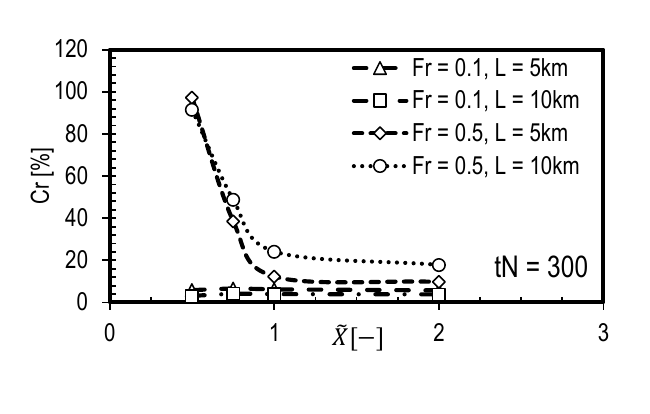}}
    \hfill
    \subcaptionbox{Reflection coefficient as a function of domain height\label{fig:wc_lz}}{\includegraphics[width=8.8cm,trim={0cm 0.5cm 0 0.5cm},clip]{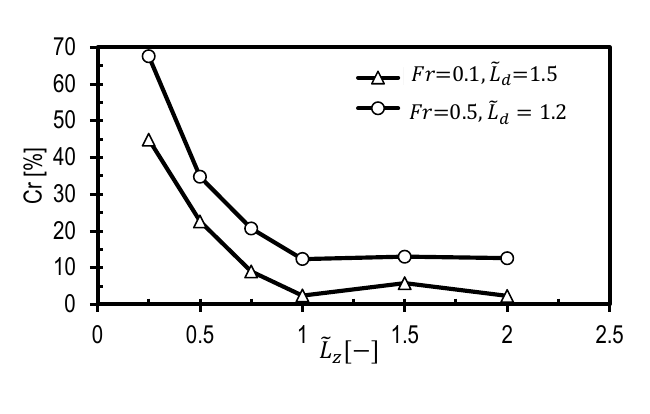}}
    \hfill\strut
    \caption{The impact of domain length and height on reflections of wind farm canopy induced IGWs.}
    \label{fig:wc_x_lz}
\end{figure}

The domain lengths were configured in simulations with the horizontal wavelength predicted from the linear theory for a hill length the same as the wind farm canopy. Therefore, in \cref{fig:wc_x}, domain length and height are normalized with predicted effective wavelengths. We calculate the effective wavelengths from the simulations at $tN = 300$ to confirm the predicted wavelengths. The horizontal wavelength is twice the length between the consecutive crest and trough with the highest amplitudes on the vertical velocity at $\tilde{L}_z = 1.0$ along the domain length. The same applies to vertical wavelength, except the vertical velocity is taken along the domain height at $X = 0$. 

It is interesting to notice that the wave shape for the canopies is significantly different from that of the hill for the same conditions and optimal simulation setup. The vertical velocity plots for two wind farm canopy lengths are shown in \cref{fig:wContours_canopy}. There are two prominent wave trains in the $Fr = 0.1$ cases, first at the entrance and the other at the canopy exit. The most dominant is caused by upward flow deflection at the entrance, stretching over most of the canopy. In contrast, the wave at the canopy exit results from the down flow as the thrust force abruptly ends. These waves are out of phase and propagate at the same angles to the horizontal. Thus, their interaction leads to a distorted wave spectrum. Therefore, the effective wavelengths from simulations with canopies differ from the predicted ones if the most dominant one is considered. The predicted and calculated wavelengths match well when the global maxima at the entrance and global minima at the exit are taken for the calculated one.  However, referring to the monotonic $Cr$ plots in \cref{wc_cr_vs_ld_fr0.1} and nearly constant values in \cref{fig:wc_x} for $Fr = 0.1$ indicates the dominant wave train to be more critical in terms of simulation setup. This wave train propagates upstream, and the other one superimposes onto it if it propagates upstream. For $Fr = 0.5$, we see only one wave train because the advection speeds, $25$ and $50~ms^{-1}$, completely dominate the wave speed. These advection speeds are at the higher end and beyond the cut-out wind speeds but are considered here to maintain $Fr = 0.5$ for practical values of $N$ when $L$ is a constraint. 

Another important observation is the enhanced wave energy 
accumulation of at the inlet in all cases, suggesting that Dirichlet inflow boundary conditions are inappropriate to simulate AGWs. Opposed to the zero gradient outflow boundary condition, which is aware of the flow reaching it, the inflow boundary condition is unaware of approaching waves. Thus, it feeds a user-prescribed inflow velocity that is a misrepresentation of the flow with AGWs in it. As a consequence, the wave energy accumulates at the inlet, and the inlet-RDL can only delay its propagation back into the domain. This requires a separate investigation of methods to contain or eliminate the energy accumulation at the inlet. 

Going back to the context of this study, the waves are effectively damped, especially in the top and outlet RDLs, suggesting that the optimal setups from the hill case are also optimal for the wind farm canopies. Quantitatively, examining reflections in \cref{fig:wc_x} shows no impact of domain length for the $Fr = 0.1$. This is because the domain lengths were configured to $\lambda_{hor}$ predicted from the linear theory, which is significantly greater than the wavelength of the dominant wave train at the entrance of the WFC. This means that space more than WFC-induced wavelengths was available in the simulation setup.  However, the decreasing reflection trend for increasing domain length remains for $Fr = 0.5$ cases; $Cr$ is the minimum for $\tilde X > 1.0$. However, $Cr$ values are more than $10\%$ because the strong amplitudes for high advection speeds and the RDLs seem to do only that much. 

The impact of domain height on the reflections is also the same as that of the hill case. As shown in \cref{fig:wc_lz}, $Cr$ converges well for domain heights greater than $\lambda_{ver}$. It is important to recall that vertical wavelength depends on $U/N$, which was consistent between the WFC and hill cases. The $Cr$ is higher for $Fr = 0.5$ because of high amplitudes due to high advection speed, $25~ms^{-1}$. Based on the above quantitative and qualitative discussions, it is suggested to set the domain length and height in wind farm simulations to at least one predicted effective horizontal and vertical wavelength, respectively. However, the link between high inflow velocities and wave amplitudes and trains should be investigated by modeling the wind farms with actuator models for an accurate representation of the wind farm flow dynamics.

\conclusions  
\label{sec:conclu}
This study aims to provide guidelines for atmospheric flow simulations that include atmospheric gravity waves by relating the involved physical and simulation parameters. The study is first carried out for a two-dimensional hill in stably stratified flow to compare the results to an analytical solution. The findings are then tested for flow through wind farms, approximated with a wind-farm canopy model. Based on recent findings in the literature, only Rayleigh damping gravity wave treatment was investigated. Therefore, the findings apply to simulation setups with Rayleigh damping layers and inflow/outflow boundary conditions solved with finite volume codes. 

Simulation time is one of the most critical parameters in simulations, including gravity waves. In all cases, longer simulation time would result in the accumulation of wave energy at the inlet boundary, and the reflections gradually become stronger. Thus, we conclude that the Rayleigh damping method attenuates the gravity waves to an extent that may not work for a diurnal simulation. Therefore, a robust technique is required to handle both the energy accumulation and reflections. 

The results regarding the configuration of the damping layers show a trade-off between the ability to properly admit gravity waves and computational resources. With periodic conditions in the lateral direction, the highest accuracy can be achieved with damping layers of a thickness exceeding the effective vertical wavelength at the inlet, top, and outlet. In case of limitations on the computational resources, a combination of damping layers of the same thickness at the inlet and top could still be reasonable.

Our test shows that for various Froude numbers, weak damping coefficients would not work even for damping thickness $2$ to $5$ fold greater than the effective vertical wavelength. Likewise, the reflections are higher for strong damping layers and might distort the solution in regions of the non-damped domain close to the damping layer. The most suitable damping coefficients are $1$ to $10$ when the damping coefficient is normalized with Brunt-Vaisala frequency. The thickness of the damping layer should be at least one effective vertical wavelength, and thicknesses exceeding $1.5$ times the effective vertical wavelength may be unnecessary.

The domain length should be scaled with the effective horizontal wavelength and not with the length of the terrain or wind farm. The reflection trend for domain length normalized with effective horizontal wavelength shows that large domains are more appropriate to avoid intense reflection of gravity waves from the domain boundaries. For domain lengths exceeding one horizontal wavelength, the reflection of the upward propagating energy is less than 6\% in the hill case. The reflections were slightly higher for the wind farm canopies due to the complex interaction of the wave trains at the canopy entrance and exit. Therefore, more investigation with at least the actuator disk approach to model the wind turbines is required to thoroughly test the optimal setups and understand the wave dynamics.  For both flow scenarios, increasing the domain length beyond $\lambda_{hor}$ shows a small reduction in the reflection coefficient. Similar impacts on reflection are observed for varying the non-damped domain height, and setting it to at least one effective vertical wavelength is recommended.

These recommendations are based on linearly stratified atmospheric flows. Aspects like turbulence and the complete temperature structure of the atmosphere are not considered yet. Still, this baseline can serve as a foundation for setting up simulations that include atmospheric gravity waves in wind applications. We have tested these findings with another finite volume code, TOSCA \citep{StipaEtAl2023a}, to find that they almost perfectly match the results from SOWFA.  We are exploring the impact of the inversion layer(inversion Froude number and height) on the setup. We will shortly include turbulence and ground stability (neutral, unstable, and stable) in the simulation setup to achieve the final guidelines for wind farm simulations. More importantly, we aim to come up with an inflow boundary condition that can admit the presence of AGWs and avoid the energy accumulation at the inlet. The findings related to these aspects will be reported in follow-up publications.


\dataavailability{The raw data from LES and post-processing scripts will be archived as a repository on the 4TU drive of TU Delft and will be open access.} 

\videosupplement{Videos can be made available to highlight the time evolution of the IGWs for both flow scenarios addressed in this study.} 

\appendix
\section{Configuration of Damping Layers}
\label{sec:config_dl}
The reflection of waves from the top boundary is expected as the typically used boundary conditions to model an arbitrary location in the atmosphere are mostly fully reflective. Since gravity waves travel vertically, one would anticipate reflections from the top boundary if forcing zones are not used there. Intriguingly, with inflow/outflow boundary conditions, the damping layers are needed at other boundaries, too. For instance, the reflected waves from the top direct towards the outlet, and weak reflections may also appear there. Likewise,  the wave energy in the domain distributes in the domain and accumulates over time if not damped at the boundaries. This is evident as reflections from the inlet and parallel to it propagate into the domain after one to two flow-through times. Thus, damping the waves at various boundaries is compulsory for cleaner simulations. Sometimes, the solution for a setup without an appropriate damping layer configuration will diverge due to numerical instability.  

\begin{figure}[h]
    \centering
    \includegraphics[width=0.9\textwidth]{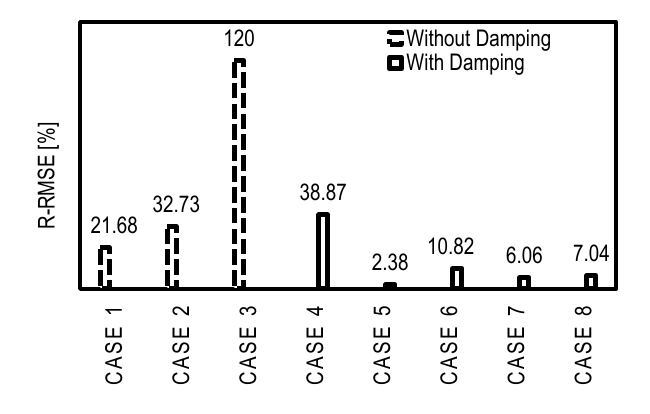}
    \caption{Comparison of R-RMSE for various damping configurations designed to acquire the most suitable arrangement for the base case.}
    \label{fig:config}
\end{figure}

In this context, deciding the configuration of damping layers becomes a critical concern while setting up simulations involving gravity waves. We explored the configuration of damping layers only to come up with a base case. The tested combinations and findings are presented in \cref{fig:config}. 

At this point, it is important to highlight that the minimum amount of reflections is a user choice. Since the reflections intensify over simulation time, it may be the case that a user will opt for a configuration based on the availability of computational resources and the simulation time of their interest. Thus, a simulation without any forcing zones is the first possibility. Interestingly, the domain height is critical when there is no damping. This can be established by comparing cases 1, 2, and 3, where the domain height ($L_z$) is 1, 2, and 3 times the effective vertical wavelength, respectively. Reflections increase rapidly for increasing the domain height, from 22\% to 33\% and 120\% in cases 1 to 3, respectively. However, none of these cases fall under the criteria for an acceptable amount of reflections. Thus, we opt to use Rayleigh damping, and having it only at the top is not enough to handle the reflections. The R-RMSE for this case is 39\%, and the solution around the hill is contaminated by the reflections from the inlet such that the actual gravity waves disappear entirely. This suggests that it is better not to have damping only at the top and instead take a damping-free setup with domain height just over one effective vertical wavelength. Indeed, this would be viable only when R-RMSE or Cr of about 20\% is acceptable for a given problem.

The most suitable case is damping layers at the inlet, outlet, and top with the same damping characteristics.  Given as case 5 in \cref{fig:config}, this configuration limits the R-RMSE to only 2.38\%. If the damping thickness is reduced by half for having damping on all three sides (Case 6), the R-RMSE increases to 10\%. A combination of damping layers at the top and outlet (Case 7) appears better than Case 6 as R-RMSE here is 6\%. Damping at the top and inlet (Case 8), is slightly more reflective (R-RMSE 7\%) than in Case 7. There can be other possible configurations, but this analysis gives enough insight to decide on a suitable configuration for a numerically stable base case. Altogether, the configuration of damping layers is a trade-off between computational resources and the desired accuracy of the solution that depends on the user's choice regarding the acceptable amount of reflections.

\noappendix       




\appendixfigures  

\appendixtables   


\authorcontribution{Conceptualization, M.A.K, D.A.; methodology, M.A.K, D.A; software, M.A.K; validation, M.A.K; formal analysis, M.A.K; investigation, M.A.K, D.A; computational resources, M.A.K, D.A; data curation, M.A.K; writing--original draft preparation, M.A.K; writing--review and editing, D.A, S.J.W, M.C; visualization, M.A.K; supervision, D.A, S.J.W, M.C; project administration, D.A, S.J.W; funding acquisition, D.A, S.J.W..
All authors have read and agreed to the published version of the manuscript.} 

\competinginterests{No competing interests are present.} 


\begin{acknowledgements}
This publication is part of the project: Numerical modelling of Regional-Scale Wind Farm Flow Dynamics, with project number: 2023/ENW/01454045 of the research programme: ENW which is (partly) financed by the Dutch Research Council (NWO).
\end{acknowledgements}




R


 \bibliographystyle{copernicus}
\bibliography{main}

\begin{thebibliography}{23}
\providecommand{\natexlab}[1]{#1}
\providecommand{\url}[1]{{\tt #1}}
\providecommand{\urlprefix}{URL }
\expandafter\ifx\csname urlstyle\endcsname\relax
  \providecommand{\doi}[1]{https://doi.org/\discretionary{}{}{}#1}\else
  \providecommand{\doi}{https://doi.org/\discretionary{}{}{}\begingroup
  \urlstyle{rm}\Url}\fi

\bibitem[{Allaerts(2016)}]{2016_Allaerts_phdthesis}
Allaerts, D.: Large-eddy Simulation of Wind Farms in Conventionally Neutral and
  Stable Atmospheric Boundary Layers, Ph.D. thesis, 2016.

\bibitem[{Allaerts(2022)}]{Allaerts_LBoW_-_Linear_2022}
Allaerts, D.: {LBoW - Linear Buoyancy Wave Package}, \doi{10.4121/21711227},
  2022.

\bibitem[{Allaerts and Meyers(2015)}]{Allaerts2015}
Allaerts, D. and Meyers, J.: {Large eddy simulation of a large wind-turbine
  array in a conventionally neutral atmospheric boundary layer}, Physics of
  Fluids, 27, 065\,108, \doi{10.1063/1.4922339}, 2015.

\bibitem[{Allaerts and Meyers(2017)}]{Allaerts2017}
Allaerts, D. and Meyers, J.: {Boundary-layer development and gravity waves in
  conventionally neutral wind farms}, Journal of Fluid Mechanics, 814, 95--130,
  \doi{10.1017/jfm.2017.11}, 2017.

\bibitem[{Allaerts and Meyers(2018)}]{Allaerts2018a}
Allaerts, D. and Meyers, J.: {Gravity Waves and Wind-Farm Efficiency in Neutral
  and Stable Conditions}, Boundary-Layer Meteorology, 166, 269--299,
  \doi{10.1007/s10546-017-0307-5}, 2018.

\bibitem[{Allaerts et~al.(2018)Allaerts, Broucke, {Van Lipzig}, and
  Meyers}]{Allaerts2018}
Allaerts, D., Broucke, S.~V., {Van Lipzig}, N., and Meyers, J.: {Annual impact
  of wind-farm gravity waves on the Belgian-Dutch offshore wind-farm cluster},
  Journal of Physics: Conference Series, 1037,
  \doi{10.1088/1742-6596/1037/7/072006}, 2018.

\bibitem[{Churchfield et~al.(2012{\natexlab{a}})Churchfield, Lee, and
  Moriarty}]{churchfield2012overview}
Churchfield, M., Lee, S., and Moriarty, P.: Overview of the simulator for wind
  farm application (SOWFA), National Renewable Energy Laboratory,
  2012{\natexlab{a}}.

\bibitem[{Churchfield et~al.(2012{\natexlab{b}})Churchfield, Lee, Michalakes,
  and Moriarty}]{Churchfield2012}
Churchfield, M.~J., Lee, S., Michalakes, J., and Moriarty, P.~J.: {A numerical
  study of the effects of atmospheric and wake turbulence on wind turbine
  dynamics}, Journal of Turbulence, 13, 1--32,
  \doi{10.1080/14685248.2012.668191}, 2012{\natexlab{b}}.

\bibitem[{Durran(1999)}]{Durran1999}
Durran, D.~R.: {Nonreflecting Boundary Conditions}, pp. 395--438,
  \doi{10.1007/978-1-4757-3081-4_8}, 1999.

\bibitem[{Frandsen(1992)}]{Frandsen1992}
Frandsen, S.: {On the wind speed reduction in the center of large clusters of
  wind turbines}, Journal of Wind Engineering and Industrial Aerodynamics, 39,
  251--265, \doi{10.1016/0167-6105(92)90551-K}, 1992.

\bibitem[{Inoue et~al.(2014)Inoue, Matheou, and Teixeira}]{InoueEtAl2014}
Inoue, M., Matheou, G., and Teixeira, J.: LES of a Spatially Developing
  Atmospheric Boundary Layer: Application of a Fringe Method for the
  Stratocumulus to Shallow Cumulus Cloud Transition, Monthly Weather Review,
  142, 3418 -- 3424, \doi{https://doi.org/10.1175/MWR-D-13-00400.1}, 2014.

\bibitem[{Klemp and Lilly(1978)}]{klemp-Lilly1978}
Klemp, J. and Lilly, D.: Numerical Simulation of Hydrostatic Mountain Waves, J.
  Atmos. Sci., 35, 78--107,
  \doi{10.1175/1520-0469(1978)035<0078:NSOHMW>2.0.CO;2}, 1978.

\bibitem[{Lanzilao and Meyers(2021)}]{Lanzilao2021}
Lanzilao, L. and Meyers, J.: {Set-point optimization in wind farms to mitigate
  effects of flow blockage induced by atmospheric gravity waves}, Wind Energy
  Science, 6, 247--271, \doi{10.5194/wes-6-247-2021}, 2021.

\bibitem[{Lanzilao and Meyers(2023)}]{Lanzilao2022}
Lanzilao, L. and Meyers, J.: {An improved fringe-region technique for the
  representation of gravity waves in large-eddy simulation with application to
  wind farms}, Boundary-Layer Meteorology, 186, 567--593,
  \doi{10.1007/s10546-022-00772-z}, 2023.

\bibitem[{Maas(2022)}]{wes-2022-63}
Maas, O.: From gigawatt to multi-gigawatt wind farms: wake effects, energy
  budgets and inertial gravity waves investigated by large-eddy simulations,
  Wind Energy Science Discussions, 2022, 1--31, \doi{10.5194/wes-2022-63},
  2022.

\bibitem[{Markfort et~al.(2018)Markfort, Zhang, and Porté-Agel}]{markfort}
Markfort, C., Zhang, W., and Porté-Agel, F.: Analytical Model for Mean Flow
  and Fluxes of Momentum and Energy in Very Large Wind Farms, Boundary-Layer
  Meteorology, 166, \doi{10.1007/s10546-017-0294-6}, 2018.

\bibitem[{Peri\'c(2019)}]{Peric}
Peri\'c, R.: Minimizing undesired wave reflection at the domain boundaries in
  flow simulations with forcing zones, Master's thesis, Technische
  Universität Hamburg, 2019.

\bibitem[{Smith(2010)}]{Smith2010}
Smith, R.~B.: {Gravity wave effects on wind farm efficiency}, Wind Energy, 13,
  449--458, \doi{10.1002/WE.366}, 2010.

\bibitem[{Snyder et~al.(1985)Snyder, Thompson, Eskridge, Lawson, Castro, Lee,
  Hunt, and Ogawa}]{snyder_thompson_eskridge_lawson_castro_lee_hunt_ogawa_1985}
Snyder, W.~H., Thompson, R.~S., Eskridge, R.~E., Lawson, R.~E., Castro, I.~P.,
  Lee, J.~T., Hunt, J. C.~R., and Ogawa, Y.: The structure of strongly
  stratified flow over hills: dividing-streamline concept, Journal of Fluid
  Mechanics, 152, 249–288, \doi{10.1017/S0022112085000684}, 1985.

\bibitem[{Stipa et~al.(2023)Stipa, Ajay, Allaerts, and
  Brinkerhoff}]{StipaEtAl2023a}
Stipa, S., Ajay, A., Allaerts, D., and Brinkerhoff, J.: TOSCA - An Open-Source
  Finite-Volume LES Environment for Wind Farm Flows, Wind Energy Science, 2023.

\bibitem[{Taylor and Sarkar(2007)}]{Taylor2007}
Taylor, J.~R. and Sarkar, S.: {Internal gravity waves generated by a turbulent
  bottom Ekman layer}, Journal of Fluid Mechanics, 590, 331--354,
  \doi{10.1017/S0022112007008087}, 2007.

\bibitem[{Vosper and Ross(2020)}]{Vosper2020}
Vosper, S.~B. and Ross, A.~N.: Sampling Errors in Observed Gravity Wave
  Momentum Fluxes from Vertical and Tilted Profiles, Atmosphere, 11,
  \doi{10.3390/atmos11010057}, 2020.

\bibitem[{Wu and Port{\'{e}}-Agel(2017)}]{Wu2017}
Wu, K.~L. and Port{\'{e}}-Agel, F.: {Flow adjustment inside and around large
  finite-size wind farms}, Energies, 10, 4--9, \doi{10.3390/en10122164}, 2017.

\end{thebibliography}

\end{document}